\def\erg{\varepsilon}

\def\sigt{\sigma_{\hbox{\fiverm T}}}

\def\taupp{\tau_{\gamma\gamma}}

\font\fiverm=cmr5

               \font\sixbf=cmbx6      
\def\apj{{\it Ap. J. }}                                
\def\apjl{{\it Ap. J. Lett. }}                         
\def\apjs{{\it Ap. J. Supp. }}                         
\def\aapl{{\it Astron. Astr. (Lett.) }}                
\def\nat{{\it Nature }}                                
\def\ssr{{\it Space Sci. Rev. }}                       
\def\teq#1{$\, #1\,$}                         
\def\dover#1#2{\hbox{${{\displaystyle#1 \vphantom{(} }\over{
   \displaystyle #2 \vphantom{(} }}$}}                
{\catcode`\@=11                                                  
\gdef\SchlangeUnter#1#2{\lower2pt\vbox{\baselineskip 0pt\lineskip0pt    
\ialign{$\m@th#1\hfil##\hfil$\crcr#2\crcr\sim\crcr}}}}           
\def\gtrsim{\mathrel{\mathpalette\SchlangeUnter>}}               
\def\lesssim{\mathrel{\mathpalette\SchlangeUnter<}}    

\def\eb{\erg_{\hbox{\fiverm B}}}
\def\Eb{E_{\hbox{\fiverm B}}}

\documentstyle[psfig]{aipproc}

\begin{document}
\newcommand{\vol}[2]{$\,$\bf #1\rm , #2}                 
\vphantom{p}
\vskip -55pt
\centerline{\hfill To appear in Proc. of Snowbird TeV Gamma-Ray Workshop}
\centerline{\hfill ed. B.~L. Dingus (AIP, New York, 1999)}
\vskip 15pt

\title{High-Energy Spectral Signatures\\ in Gamma-Ray Bursts}

\author{Matthew G.~Baring$^{\dagger}$}
\address{Laboratory for High Energy Astrophysics, Code 661\\
NASA Goddard Space Flight Center, Greenbelt, MD 20771\\
{\it baring@lheavx.gsfc.nasa.gov}\\
$^{\dagger}$Universities Space Research Association}

\maketitle

\begin{abstract}
One of the principal results obtained by the EGRET experiment aboard
the Compton Gamma-Ray Observatory (CGRO) was the detection of several
$\gamma$-ray bursts (GRBs) above 100 MeV.  The broad-band spectra
obtained for these bursts gave no indication of any high energy
spectral attenuation that might preclude detection of bursts by
ground-based \v{C}erenkov telescopes (ACTs), thus motivating several
TeV observational programs.
This paper explores the expectations for the spectral properties in the
TeV and sub-TeV bands for bursts, in particular how attenuation of
photons by pair creation internal to the source modifies the spectrum
to produce distinctive spectral signatures.  The energy of spectral
breaks and the associated spectral indices provide valuable information
that can constrain the bulk Lorentz factor of the GRB outflow at a
given time.  These characteristics define palpable observational goals
for ACT programs, and strongly impact the observability of bursts in
the TeV band.
\end{abstract}

\section*{Introduction}
 \label{sec:intro}

High energy gamma-rays have been observed for six gamma-ray bursts by
the EGRET experiment on CGRO.  Most conspicuous among these
observations is the emission of an 18 GeV photon by the GRB940217 burst
\cite{hurl94}.  Taking into account the instrumental field of view,
these detections indicate that emission in the 1 MeV--10 GeV range is
probably common among bursts, if not universal.  One implication of GRB
observability at energies around or above 1 MeV is that, at these
energies, spectral attenuation by two-photon pair production
(\teq{\gamma\gamma\to e^+e^-}) is absent in the source.
From this fact, early on Schmidt \cite{Schmidt78} concluded that if
a typical burst produced quasi-isotropic radiation, it had to be less
distant than a few kpc, since the optical depth \teq{\taupp} scales as
the square of the distance to the burst.

This result conflicted with BATSE's determination of the spatial
isotropy and inhomogeneity of bursts \cite{Meeg96}, which suggested
that they are either in an extended halo or at cosmological distances
(where \teq{\taupp\sim 10^{12}} for isotropic photons).  Hence Fenimore
et al. \cite{feh92} proposed that GRB photon angular distributions are
highly beamed and produced by a relativistically moving
plasma, a suggestion that has become very popular.  This can
dramatically reduce \teq{\taupp} and blueshift spectral attenuation
turnovers out of the observed spectral range.  Various determinations
of the bulk Lorentz factor \teq{\Gamma} of the GRB medium have been
made in recent years, mostly concentrating \cite{kp91,baring93} on
cases where the angular extent of the source was of the order of
\teq{1/\Gamma}.  These calculations generally assume an infinite
power-law burst spectrum, and deduce \cite{bh97b} that gamma-ray
transparency up to the maximum energy detected by EGRET requires
\teq{\Gamma\gtrsim 100}--\teq{10^3} for cosmological bursts.

While power-law source spectrum assumption is expedient, the spectral
curvature seen in most GRBs by BATSE \cite{band93} is expected to play
an important role in reducing the opacity for potential TeV emission
from these sources (Baring \& Harding \cite{bh97a}).  Such curvature is
patently evident in 200 keV--2 MeV spectra of some EGRET-detected bursts,
and its prevalence in bursts is indicated by the generally steep EGRET
spectra for bursts \cite{hurl94,Schneid92,Sommer94}.  In this
paper, the principal effects introduced into pair production opacity
calculations by spectral breaks in the BATSE energy range are
considered, focusing the work of \cite{bh97a} to identify the
properties of cosmological bursts in the 1 GeV--1 TeV range.  These
signatures are clearly distinguishable from absorption by background
radiation fields, thereby defining diagnostics that future ground-based
initiatives such as Veritas, MILAGRO, HESS and MAGIC, and space
missions such as GLAST can provide for GRB studies.

\section*{SPECTRAL CURVATURE AND $\gamma$-$\gamma$ ATTENUATION}
                                                 
The simplest picture \cite{kp91,baring93} of relativistic beaming has
``blobs'' of material moving with a bulk Lorentz factor \teq{\Gamma}
more-or-less toward the observer, and having an angular ``extent''
\teq{\sim 1/\Gamma}.  For an infinite power-law spectrum \teq{n(\erg
)=n_{\gamma} \,\erg^{-\alpha }}, where \teq{\erg} is the photon energy
in units of \teq{m_ec^2} (a dimensionless convention used throughout),
for which the optical depth to pair creation assumes the form
\teq{\taupp (\erg )\propto \erg^{\alpha -1}\Gamma^{-(1+2\alpha )}} for
\teq{\Gamma\gg 1}.  As noted above, the input source spectrum needs to
be modified, to explore the effects of a relative depletion of low energy
photons in the BATSE range.  The simplest approximation to spectral
curvature is a power-law broken at a dimensionless energy \teq{\eb}
(\teq{=\Eb /0.511}MeV):
\begin{equation}
   n(\erg )\; =\; n_{\gamma}\eb^{-\alpha_h}\, \cases{
   \eb^{\alpha_l}\erg^{-\alpha_l} , &
     if \teq{\vphantom{\Bigl(} \erg\leq\eb\;\;},\cr
   \eb^{\alpha_h}\erg^{-\alpha_h} , & \teq{\erg >\eb\;\; }. \cr }
 \label{eq:powerlaw}
\end{equation}
The optical depth determination for such a distribution utilizes
results obtained in \cite{gs67} for truncated power-laws.  The
resulting forms are presented in \cite{bh97a}, and the optical depth
\teq{\taupp (\erg )} for attenuation of a broken power-law photon
distribution has the basic form
\begin{equation}
   \dover{\taupp (\erg )}{n_{\gamma}\sigt R}\;\propto \; \cases{
   \dover{\erg^{\alpha_h-1}}{\Gamma^{2\alpha_h}}, &
      if \teq{\vphantom{\Bigl(} \erg\lesssim\Gamma^2/\eb\;\;},\cr
   \dover{\erg^{\alpha_l-1}}{\Gamma^{2\alpha_l}}, &
      if \teq{\vphantom{\Bigl(} \erg\gtrsim\Gamma^2/\eb\;\;},\cr }
 \label{eq:tauppform}
\end{equation}
that implies breaks in the absorbed portion of the hard
gamma-ray spectrum that ``image'' the BATSE band break in the seed
photons.  More gradual spectral curvature can be treated by
fitting the GRB continuum with piecewise continuous power-laws.  A
variability ``size'' \teq{R_v=3\times 10^7}cm (\teq{=R/\Gamma}) is
chosen here following \cite{bh97b,bh97a}, and the observed flux at 1
MeV normalizes the source density coefficient \teq{n_{\gamma}}.

\begin{figure}[t]
\centerline{\psfig{figure=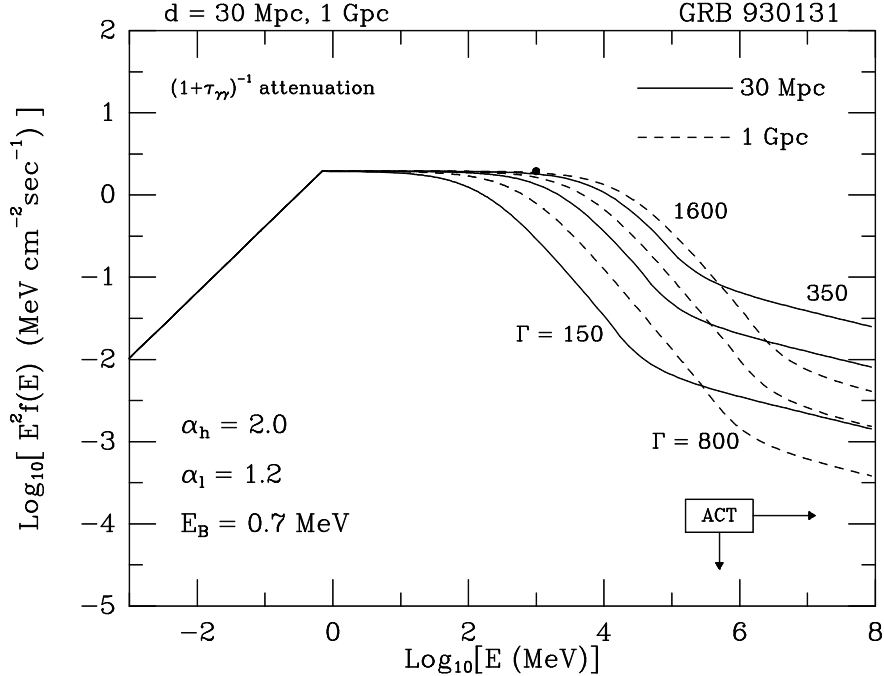,height=8.9truecm}}
\caption{The $\gamma$-$\gamma$ attenuation, internal to the source,
for GRB 930131 at distances typical of nearby (solid curves,
\teq{\Gamma =150,\, 250,\, 350}) and distant (short dashed curves,
\teq{\Gamma =800,\, 1200,\, 1600}) cosmological origin, for bulk
Lorentz factors \teq{\Gamma} of the emitting region.  The source
spectrum (\teq{\nu F_{\nu}} format) was a power-law broken at
\teq{E_{\rm B} =0.7}MeV, with spectral indices \teq{\alpha_l=1.2} and
\teq{\alpha_h=2.0}.  The filled circle denotes the highest energy EGRET
photon at 1000 MeV [7].  The threshold and sensitivity for ACT
observations of later bursts is indicated by the ``ACT'' box.
 \label{fig:atten30Mpc1Gpc}}
\end{figure}

The results of the attenuation of the spectra in
Eq.~(\ref{eq:powerlaw}) are depicted in Fig.~\ref{fig:atten30Mpc1Gpc}
using an attenuation factor \teq{1/(1+\taupp )} that is appropriate for
opacity skin depth effects.  The source spectrum parameters are chosen
to approximate the observed values for the ``Superbowl burst''
GRB930131, for two different extragalactic distance scales: the nearer
one, 30 Mpc, is appropriate to scenarios where GRBs generate ultra-high
energy cosmic rays.  Clearly the attenuation is marked in the GeV--TeV
band for the Lorentz factors \teq{\Gamma} chosen, and could be reduced
by increasing \teq{\Gamma}.  The onset of attenuation couples to
\teq{\Gamma} and the EGRET band spectral index \teq{\alpha_h}, and
above this turnover the immediate spectral index is \teq{1-2\alpha_h}.
Precise knowledge of the GRB distance, such as through redshifts of
accompanying optical afterglows, would facilitate the determination of
tight constraints on \teq{\Gamma}.  For either distance scale in
Fig.~\ref{fig:atten30Mpc1Gpc}, there is a flattening (to index
\teq{1-\alpha_h-\alpha_l}) in the TeV/sub-TeV band that is a
consequence of the spectral break in the BATSE band:  it arises at
\teq{\erg\sim\Gamma^2/\eb}.


\begin{figure}[t]
\centerline{\psfig{figure=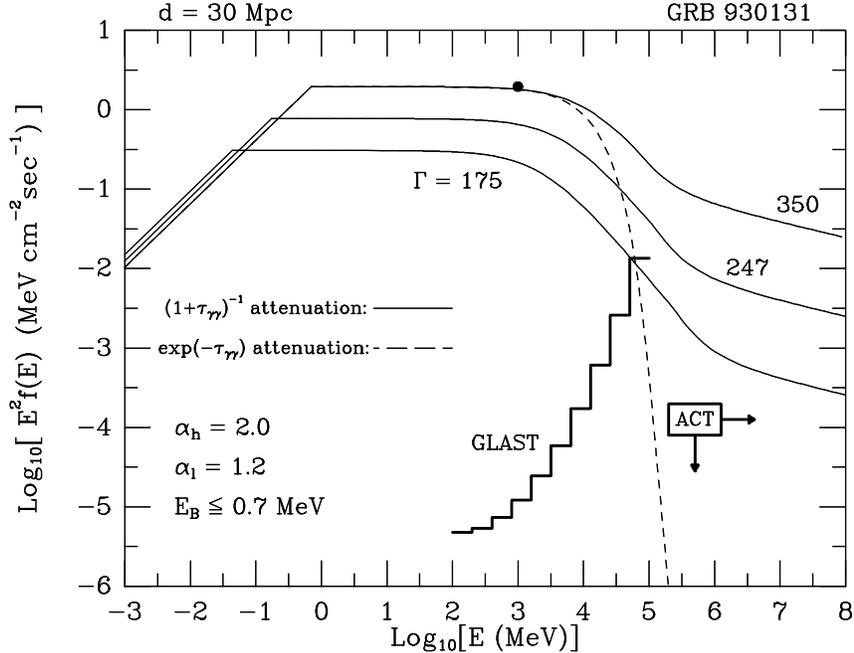,height=8.6truecm}}
\caption{An evolutionary sequence for \teq{\gamma}-\teq{\gamma}
attenuation, starting with a GRB930131 data fit, appropriate to an
adiabatically-decelerating blast wave.  
The GLAST steady-source differential
(\teq{\Delta E/E=2} step-function) and approximate Whipple integral
sensitivities (ACT box; derived from upper limits to later bursts) are
depicted.  A case of exponential attenuation (for the \teq{\Gamma=350},
dashed line) is illustrated; it would inhibit detections by ACTs.
 \label{fig:atten_timedep}}
\end{figure}

The potential for observational diagnostics is immediately apparent.
First, the extant EGRET data already provides a lower bound to
\teq{\Gamma}: the dot on Fig.~\ref{fig:atten30Mpc1Gpc} represents the
highest energy photon from GRB930131, and clearly suggests that
\teq{\Gamma\gtrsim 250} for \teq{d=30}Mpc or \teq{\Gamma\gtrsim 800}
for \teq{d=1}Gpc.  Second, the sensitivity of ACTs is easily sufficient
to detect bursts even with significant attenuation, so that they could
well probe the spectral issues raised here.  While the Whipple rapid
search \cite{Conn97} postdated the EGRET detections, and produced
merely upper limits as indicated in Fig.~\ref{fig:atten30Mpc1Gpc}, an
intriguing possible detection of a BATSE burst by the MILAGRITO
forerunner to MILAGRO was announced at this meeting by McEnery et al.
(these proceedings), foreshadowing advances to come.

Perhaps the greatest strides in understanding will be precipitated by
broad-band spectral coverage afforded by simultaneous detection of
bursts by GLAST and TeV experiments like MILAGRO.
Fig.~\ref{fig:atten_timedep} displays a time-evolutionary sequence of
GRB spectra, including the effects of \teq{\gamma}-\teq{\gamma}
attenuation, and compares this with the potentially-constraining
current Whipple integral sensitivity threshold (deduced from the
results of \cite{Conn97}), and the projected GLAST {\it
steady-source differential} sensitivity.  The GLAST sensitivity is
obtained from simulations (Digel, private communication) of the
spectral capability for high latitude, steady sources in a one-year
survey, i.e. roughly 8 weeks on source.  The real GLAST sensitivity for
transient GRBs of duration \teq{t_{\rm dur}} can be estimated to be
roughly \teq{[(8\, {\rm weeks})/t_{\rm dur}]^{1/2}} times that depicted.
Note that the differential sensitivity is the most appropriate measure
for spectral diagnostic capabilities.  Evidently, ACTs and GLAST
working in concert will be able to determine the spectral shape and
evolution of bright, flat-spectrum bursts like GRB930131 if the
attenuation is no more dramatic than \teq{1/(1+\taupp )}.  The
particular evolutionary scenario depicted in
Fig.~\ref{fig:atten_timedep} is an adiabatic one for blast wave
deceleration during the sweep-up phase, where the dependences on time
\teq{t} are \teq{\Gamma\propto t^{-3/8}}, \teq{\eb\propto
\Gamma^4\propto t^{-3/2}}, and \teq{\eb^2 f(\eb )\propto \Gamma^{8/3}\propto
t^{-1}} for the flux at the peak \cite{Derm99}.  Shifts in the
turnover energy and sub-TeV break energy, and correlations with BATSE
flux and break energy should be discernible in bright sources.

It must be emphasized that these internal absorption characteristics
are easily distinguishable from those of external absorption due to the
cosmological infra-red background along the line of sight
\cite{SdeJ96,mhf96}.  Attenuation by such background fields couples to
the redshift, not parameters internal to the source nor the shape of
the spectrum in the BATSE and EGRET bands.  Furthermore, it is always
exponential in nature (i.e. of severity equivalent to the dashed curve
in Fig.~\ref{fig:atten_timedep}) since the emission region is distinct
from the location of the soft target photons, and is patently
independent of time.  The possibility of confusing such with the
internal attenuation that forms the focus of this paper seems minimal.
Hence, the prospects for powerful spectral diagnostics in bright bursts
with atmospheric \v{C}erenkov telescopes and the GLAST mission promise
an exciting time ahead for the field of high energy gamma-ray
astronomy.

\vskip 5pt\noindent
{\bf Acknowledgments:}  I thank Alice Harding and Brenda Dingus for
helpful discussions, and Seth Digel for simulating GLAST spectral
sensitivities.

\end{document}